\documentstyle[12pt,epsfig]{article}
\newcommand{\beq}{\begin{equation}}
\newcommand{\eeq}{\end{equation}}
\newcommand{\beqn}{\begin{eqnarray}}
\newcommand{\eeqn}{\end{eqnarray}}
\newcommand{\beqnn}{\begin{eqnarray*}}
\newcommand{\eeqnn}{\end{eqnarray*}}

\newcommand{\MT}{M_\perp}
\newcommand{\QT}{Q_\perp}

\newcommand{\EllE}{{\rm E}\hspace*{0.1em}}
\newcommand{\EllF}{{\rm F}\hspace*{0.1em}}

\newcommand{\II}{{\rm I}\hspace*{0.1em}}

\begin{document}
\thispagestyle{empty}

\vspace*{2cm}

\begin{center}
{\large \bf
Can one  discriminate the thermal dilepton signal against
the open charm and bottom decay background in ultrarelativistic
heavy-ion collisions?} \\[6mm]
{\sc
K. Gallmeister$^a$, B. K\"ampfer$^a$, O.P. Pavlenko$^{a,b}$} \\[6mm]

$^a$Forschungszentrum Rossendorf, PF 510119, 01314 Dresden, Germany \\[1mm]
$^b$Institute for Theoretical Physics, 252143 Kiev - 143, Ukraine
\end{center}

\vspace*{9mm}

\centerline{{\bf Abstract}} 
In ultrarelativistic heavy-ion collisions
at $\sqrt{s} >$ 20 (120) A$\cdot$GeV a copious production of charm (bottom)
production sets in which,
via correlated semileptonic $D \bar D$ ($B \bar B$) decays, gives rise
to a dilepton yield at invariant mass $M \approx$ 2 -- 3  GeV in excess
of the Drell-Yan yield and the thermal dilepton signal from
deconfined matter as well.
We show that appropriate single-electron transverse momentum cuts
(suitable for ALICE at LHC) cause a threshold like behavior
of the dilepton spectra from heavy-quark meson decays 
and the Drell-Yan process and can allow
to observe a thermal dilepton signal from hot deconfined matter.\\[3mm]
{\it PACS number(s):\/}
25.75.+r, 12.38.Mh, 24.85.+p\\[2cm]
\newpage
\section{Introduction} 

The current relativistic heavy-ion experiments at CERN-SPS and planned
future experiments at BNL-RHIC and CERN-LHC are aimed at a study of the
properties of highly excited, deconfined matter.
The production of dileptons with intermediate invariant mass in the so-called
continuum region between $\phi$ and $J/\psi$ are widely considered
as one of the main tools to measure in such experiments the ''initial''
temperature and other thermodynamical characteristics of the produced matter
\cite{Shur3,Ruusk}. Dilepton measurements are envisaged in particular
with the PHENIX and ALICE detector facilities at RHIC and LHC,
respectively.
The thermal dilepton signal in the intermediate invariant mass region
faces a serious background problem connected with hard initial
quark - anti-quark annihilation into dileptons (i.e. the Drell-Yan
production). The general expectation is that with
increasing beam energy the maximum temperature of matter rises
and consequently
the thermal dilepton yield grows faster than the Drell-Yan yield.
However with increasing beam energy also a copious production of
heavy quarks, resulting in hard initial collisions of partons too,
sets in. As a consequence, the correlated semileptonic decays of open
charm and bottom mesons yield a dilepton rate exceeding the one from
the Drell-Yan process. This can be understood by the following simple
arguments. Drell-Yan pairs are produced dominantly via
$q \bar q$ annihilation, while the open charm and bottom production involves
mainly gluons of colliding nuclei. Since the parton structure functions
\cite{MRS} point to a strong increase of the gluon density at small
values of the below explained variable $x$,
one can expect a corresponding increase of the relative
contribution from charm and bottom decays into the intermediate mass
region of dileptons with increasing beam energy or $\sqrt{s}$.
Therefore it is {\it a priori} not obvious in which energy region one meets
best conditions for observing a thermal dilepton signal from
hot deconfined matter.

In this note we present a schematic view on the beam energy dependence
of the mentioned sources of dileptons. Additionally we study
systematically the influence
of single-electron transverse momentum cuts
on the invariant mass spectra of dileptons
and show that at ALICE at LHC such cuts can suppress the Drell-Yan and
correlated semileptonic decay dielectrons. Thus an
observation of the thermal signal seems to be feasible, 
supposed the uncorrelated background can be accurately enough
removed by like-sign subtraction.

\section{Beam energy dependence of dilepton sources} 

Our basic equations for calculating the above mentioned dilepton sources
can be found in Refs.~\cite{PLB97,PLB98} and are presented below in modified
form. In Fig.~1 we show the beam energy dependence of the Drell-Yan yield
and the yield from correlated $D \bar D$, $B \bar B$ decays
in central Au + Au collisions.
One infers from Fig.~1 that the ratio of Drell-Yan to open charm (bottom)
decay dileptons with invariant mass $M =$ 2.5 GeV drops down with increasing
$\sqrt{s}$. Only in the region $\sqrt{s} <$ 20 (120) A$\cdot$GeV the
Drell-Yan yield dominates above the charm (bottom) decay contribution.
Such a conclusion is in agreement with a recent analysis \cite{PBM}
which points out that the open charm source cannot explain the
experimental data of CERES and NA38/NA50 of the dilepton spectra in
S -- U and Pb -- Pb collisions at SPS energies.

Also displayed in Fig.~1 are rough estimates of the thermal signal from
purely deconfined matter assuming boost-invariant expansion.
The initial parameters for RHIC and LHC energies are
described below. For SPS energies we take the initial temperature
$T_i =$ 200 MeV at initial time $\tau_i =$ 1 fm/c, which are compatible
with the observed final pion rapidity densities.
Note that the thermal yield estimates depend sensitively on the assumed
initial parameters.
In agreement with recent findings the thermal dilepton
signal is up to two orders of magnitude below the correlated open charm
decay dileptons in a wide range of invariant masses at LHC,
RHIC \cite{Vogt2} and SPS energies.
Therefore, there is nowhere a preferred energy region;
at high beam energies, however, one expects clearer deconfinement effects
due to temperatures far above the confinement temperature.

The results displayed in Fig.~1 cover
the full phase space. Any detector acceptance will suppress the
various sources differently \cite{PRC98}. Energy losses of heavy quarks
propagating through deconfined matter \cite{Baier}
reduce also the open charm and bottom
decay yields above $M = 2$ GeV \cite{PLB98,Shur1,Lin2}.
Notice that with increasing invariant mass the thermal yield drops
exponentially while, for instance, the Drell-Yan yields drops less, i.e.
like a power law.

\section{Transverse momentum cuts} 

From all of these considerations the urgent problem arises whether one
can find such kinematical gates which enable one to discriminate the thermal
signal against the large decay background. Since the kinematics of
heavy meson production and decay differs from that of thermal dileptons,
one can expect that special kinematical restrictions superimposed on
the detector acceptance will be useful for finding the needed window for
observing thermal dileptons in the intermediate mass continuum region.
In particular, experimental cuts on the rapidity gap between the leptons
can reduce considerably the charm decay background \cite{Vogt2}.
As demonstrated recently \cite{PRC98}, the measurement of double
differential dilepton spectra as a function of the transverse pair
momentum $Q_\perp$ and transverse mass $M_\perp = \sqrt{M^2 + Q_\perp^2}$
within a narrow interval of $M_\perp$ also offers the chance to observe
thermal dileptons at LHC.
In the present note we show that a large enough low-$p_\perp$ cut on
single electrons suppresses the mentioned background processes and
opens a window for the thermal signal
in the invariant mass distribution. We take into account
the acceptance of the ALICE detector at LHC: the single electron
pseudo-rapidity is limited to $\vert \eta_e \vert \le 0.9$ and an
overall low-$p_\perp$ cut of 1 GeV is applied (we do not impose
the earlier planned restriction $p_\perp < 2.5$ GeV).
We are going to study systematically the effect of enlarging the
low-$p_\perp$ cut on single electrons in the invariant-mass
dilepton spectra at $\sqrt{s} =$ 5500 GeV.

Since the energy of individual decay electrons (positrons) has a
maximum of about 0.88 (2.2) GeV in the rest frame of the decaying
$D$ ($B$) meson, one can expect to get a strong suppression of correlated
decay lepton pairs by choosing a high enough low-momentum cut
$p_\perp^{\rm min}$ on the individual leptons in the mid-rapidity region.
For thermal leptons stemming from deconfined matter there is no such upper
energy limit and for high temperature the thermal yield will not suffer
such a drastically suppression by the $p_\perp^{\rm min}$ cut as the
decay background.
To quantify this effect we perform Monte Carlo calculations generating
dileptons from the above mentioned sources. We first employ the
transparent lowest-order calculations and then present 
more complete PYTHIA simulations. 

\subsection{Dilepton sources}
\subsubsection{Heavy quark production and decay}

We employ here the leading order QCD processes $gg \to Q \bar Q$
and $q \bar q \to Q \bar Q$ for heavy quark production and simulate
higher order corrections by an appropriate constant K factor.
The number of $Q \bar Q$, produced initially with
momenta $p_{\perp 1} = - p_{\perp 2} = p_\perp$ at rapidities $y_{1,2}$
in central $AA$ collisions can be calculated by
\begin{eqnarray} \label{eq.1}
dN_{Q \bar Q}
& = &
T_{AA}(0) \, {\cal K}_Q \, H(y_1,y_2,p_\perp) \,
dp_\perp^2 \, dy_1 \, dy_2, \\
H(y_1,y_2,p_\perp)
& = &
x_1 \, x_2 \left\{ f_g(x_1,\hat Q^2) \, f_g(x_2,\hat Q^2)
\frac{d \hat \sigma_g^Q}{d \hat t} \right. \nonumber \\
&& +
\left. \sum_{q \bar q} \left[
f_q(x_1,\hat Q^2) \, f_{\bar q} (x_2,\hat Q^2) +
f_q(x_2,\hat Q^2) \, f_{\bar q} (x_1,\hat Q^2)\right]
\frac{d \hat \sigma_q^Q}{d \hat t} \right\},
\end{eqnarray}
where $\hat \sigma^Q_{q,g}/d \hat t$ are
elementary cross sections (see for details \cite{PLB97,Vogt2}),
$f_{g,q,\bar q} (x,\hat Q^2)$
denote the parton structure functions,
$x_{1,2} = m_\perp \left(
\exp\{ \pm y_1 \} + \exp\{ \pm y_2 \}  \right) /\sqrt{s}$ and
$m_\perp = \sqrt{p_\perp^2 + m_Q^2}$.
As heavy quark masses we take $m_c =$ 1.5 GeV and $m_b =$ 4.5 GeV.
We employ the HERA supported structure function set MRS D-' \cite{MRS}
from the PDFLIB at CERN.
Nuclear shadowing effects are not needed to be included for our present
order of magnitude estimates.
The overlap function for central collisions is
$T_{AA}(0) = A^2/(\pi R_A^2)$
with  $R_A = 1.2 A^{1/3}$ fm and $A = 200$.
From a comparison with results of Ref.~\cite{Vogt2}
we find the scale $\hat Q^2 = 4 m_Q^2$ and ${\cal K}_Q =$ 2 as most
appropriate.
We employ a delta function fragmentation scheme for the heavy quark
conversion into $D$ and $B$ mesons.
Dilepton spectra from correlated decays,
i.e., $D (B) \bar D (\bar B) \to e^+ X e^- \bar X$,
are obtained from a Monte Carlo code which utilizes the inclusive
primary electron energy distribution as delivered by
JETSET~7.4. The heavy mesons are randomly decayed in their rest system
and the resulting electrons then boosted appropriately.
The average branching ratio of $D (B) \to e^+ X$
is taken as 12 (10)\% \cite{PDB}. 

\subsubsection{Drell-Yan process}

Similar to heavy quarks, our calculations of the Drell-Yan pairs
are based on the lowest order $q \bar q \to e^+ e^-$ process with
appropriate K factor.
For the Drell-Yan production process of leptons at rapidities $y_+$ and
$y_-$ and transverse momenta $p_{\perp +} = - p_{\perp -} = p_\perp$
we have \cite{PLB98}
\begin{eqnarray} \label{eq.1l}
dN_{l \bar l}^{DY} & = & T_{AA}(0) \, {\cal K}_{DY} \, L(y_+,y_-,p_\perp) \,
dp_\perp^2 \, dy_+ \, dy_-, \\
L(y_+,y_-,p_\perp) & = & \sum_{q,\bar q} x_1 x_2
\left[
f_q(x_1,\hat Q^2) \, f_{\bar q} (x_2,\hat Q^2) +
f_q(x_2,\hat Q^2) \, f_{\bar q} (x_1,\hat Q^2) \right]
\frac{d \hat \sigma_q^{l \bar l}}{d \hat t},
\end{eqnarray}
with
$d \hat \sigma_q^{l \bar l} / d \hat t = \frac{\pi \alpha^2}{3}
\mbox{ch} (y_+ - y_-) / \left( 2 p_\perp^4
[1 + \mbox{ch}(y_+ - y_-)]^3 \right)$,
$x_{1,2} = p_\perp \left(
\exp\{ \pm y_+ \} + \exp\{ \pm y_- \} \right)$ $/\sqrt{s}$,
$\alpha = 1/137$,
$\hat Q^2 = x_1 x_2 s = M^2$
and ${\cal K}_{DY} =$ 1.1.\\

\subsubsection{Thermal yield}

In case of thermal emission of dileptons from deconfined matter
the needed distribution of leptons with respect to rapidities
$y_{\pm}$ and transverse momenta $p_{\perp \pm}$ reads \cite{PRC98}
\begin{equation}
dN
=
\frac{\alpha^2 R_A^2}{4\pi^4}F_q
\int d\tau \tau
\ K_0\left(\frac{\MT}{T}\right) \, \lambda_q^2 \,
d^2p_{\perp +} d^2p_{\perp -} dy_+ dy_-
\label{gen_exp}
\end{equation}
with $F_q = \sum_q e_q^2 = \frac23$ for u,d,s quarks,
$K_n$ as modified Bessel function of $n$th order, and
with dilepton transverse mass
$M_\perp^2 = p_{\perp+}^2 + p_{\perp-}^2 +
2 p_{\perp+} p_{\perp-} \mbox{ch} (y_+ - y_-)$.
The integration is to be performed on the proper time $\tau$
of the longitudinally
expanding deconfined matter with temperature $T(\tau)$ and quark fugacity
$\lambda (\tau)$ \cite{PLB97}.
Our choice of initial conditions for produced deconfined matter is based
on the estimates of Refs.~\cite{PLB97,Eskola} for the mini-jet plasma
which are similar to the self-screened parton cascade model \cite{E.M.W.}.
We take as main set of parameters the initial temperature
$T_i =$ 1000 MeV, gluon fugacity
$\lambda_i^g =$ 0.5, and light quark fugacity
$\lambda_i^q = \frac 15 \lambda_i^g$
of the parton plasma formed at LHC at initial time
$\tau_i =$ 0.2 fm/c. For the sake of definiteness we assume
full saturation at confinement temperature $T_c =$ 170 MeV
and a quadratic time dependence of $\lambda^{q,g}(\tau)$
according to the studies \cite{PLB97,PRC95}.

\subsection{Results of lowest-order calculations}

The results of our lowest-order calculations of the invariant mass spectrum
for various values of $p_\perp^{\rm min}$ are displayed in Figs.~2 -- 4.
Comparing Figs.~2 and 3, one observes that already the cut
$p_\perp^{\rm min} =$ 2 GeV causes a strong suppression of the correlated
charm decay and Drell-Yan background in the region $M \le 2 p_\perp^{\rm min}$.
The above value of the invariant mass threshold can be estimated by using
the relation
$M^2 =2 p_{\perp+} p_{\perp-} [\mbox{ch} (y_+ -y_-) - \cos(\phi_+ - \phi_-)]$,
where $\phi_{\pm}$ denote the azimuthal angles of the leptons
in the transverse plane.
In order to exceed the cut $p_\perp^{\rm min}$ most easily, the decay
leptons should go parallel to the parent heavy mesons, which in turn
are back-to-back (in the transverse plane)
in lowest order processes. As a consequence,
$\cos(\phi_1 - \phi_2) \approx -1$ and 
the minimum invariant mass becomes 
$M^{\rm min} \approx 2 p_\perp^{\rm min}$
for such decay pairs. For correlated bottom decay the electron energy
is larger in the meson rest system and both leptons can easily overcome
the threshold $p_\perp^{\rm min}$ without such strong back-to-back correlation.
Selecting, however, electrons with $p_\perp > p_\perp^{\rm min} =$ 3 GeV one can
also get the corresponding threshold like behavior for the lepton pairs
from correlated bottom decay, see Fig.~4.
Therefore the thermal signal becomes clearly visible for such a value
of $p_\perp^{\rm min}$ due to the strong suppression of the considered
background channels.

The threshold behavior does not change if we include in our calculations
energy loss effects of heavy quarks in deconfined matter \cite{Baier}.
Such effects cause mainly a suppression of the
decay contributions (cf. \cite{PLB98,Shur1,Lin2}).
We also mention
that shadowing effects, not included in our Drell-Yan and heavy quark
production estimates, will diminish these yields somewhat. (In the
mini-jet estimates of our initial conditions \cite{PLB97} shadowing effects
are already included).

\subsection{Results from PYTHIA}

The lowest order Drell-Yan yield has anyway $M^{\rm min} = 2 p_\perp^{\rm min}$.
From the above given relations for $M$ and $M_\perp$ one can derive the
inequality
$M \ge 2 p_\perp^{\rm min}
\sqrt{1 - \left(\frac{Q_\perp}{2 p_\perp^{\rm min}}\right)^2}$.
This relation tells us that in the region $M <$ 5 (4) GeV only pairs with
total transverse momentum $Q_\perp >$ 3.25 (4.5) can contribute if
$p_\perp^{min} =$ 3 GeV.
Since the next-to-leading order Drell-Yan distribution
$dN/dM^2 dQ_\perp^2 dY$ drops from $Q_\perp \approx$ 0 to 3 - 4 GeV
by nearly three orders of magnitude \cite{HardProbes}
one can estimate a small
higher order Drell-Yan contribution in the small-M region.
To quantify the smearing of the threshold effect by an intrinsic
$p_\perp$ distribution of initial partons we perform simulations with
the event generator
PYTHIA (version 6.104 \cite{PYTHIA}) with default switches.
Results are diplayed in Fig.~5 and
show that the sharp threshold
effect from the above lowest-order Drell-Yan process is indeed
somewhat smeared out,
however, the small-$M$ region is still clean.
It turns out that the initial state radiation of partons
before suffering a hard collision is the main reason for rising the
pair $Q_\perp$ and for smearing out the sharp threshold effect, while
the intrinsic $p_\perp$ distribution of partons
causes minor effects.

Let us now consider heavy quark pairs. Here,
the intrinsic $p_\perp$ distribution, and both initial and
final state radiations of the partons can cause a finite $Q_\perp$ and
thus destroy the strong back-to-back correlation,
i.e. $p_{\perp 1} \ne p_{\perp 2}$.
In Fig.~6 we show results of simulations with PYTHIA for the bottom
channel. The resulting primary dileptons from all correlated open bottom
mesons are displayed. One observes that the initial state radiation
causes a pronounced smearing of the threshold effect discussed above.
Without the initial state radiation, as implemented in PYTHIA, the
threshold effect is recovered. The intrinsic $p_\perp$ distribution
and final state radiation are negligible. The conclusion of such
studies is that an enlarged value of $p_\perp^{\rm min}$ 
is necessary to keep
clean the low-$M$ region from open bottom decay products. For charm
the smearing effect due to initial state radiation in PYTHIA is
efficiently suppressed by the large enough low-$p_\perp$ cut of 
$p_\perp^{\rm min} =$ 3 GeV.
The different behavior of charm and bottom stem from the fact that the
bottom-$p_\perp$ distribution is much wider. As a consequence, the
bottom is evolved in the average to much larger values of $\hat Q^2$
thus experiencing stronger kicks by initial state radiation.
In agreement with our previous findings \cite{PLB98}, bottom therefore
causes the most severe background processes at LHC energies.

With PYTHIA also the dileptons from single decay chain of open bottom
are accessible. This channel provides a contribution peaking at 1.5
GeV;
the cut $p_\perp^{\rm min} = 3$ GeV pushes all invariant masses below
3 GeV. Therefore, unless enlarging $p_\perp^{\rm min}$ considerably,
it will be difficult to suppress kinematically the background below the
thermal signal at invariant masses $M < 3$ GeV. Probably explicit
identification and subtraction of the bottom contribution is needed
in this region. Otherwise one should stress that the thermal signal of
deconfined matter is expected to exceed the thermal hadron signal at
$M > 2$ GeV \cite{PRC95}. Hence the region of small values of $M$ is
not interesting in this respect. 

We conclude this subsection by mentioning that further differences
between our above lowest-order calculations and the PYTHIA simulations
stem from slightly different charm/bottom masses, different fragmentation
schemes and decay chains. We have checked that for the same masses, the
same fragmentation scheme (i.e. independent fragmentation) and PYTHIA
without intrinsic $p_\perp$ distribution and without initial/final
state radiation, PYTHIA then reproduces our calculations. According to
our experience, the stringent kinematical cuts amplify small
differences in various code versions. We do not intent here to present
a detailed prediction of the dilepton yield, but rather to demonstrate
that an observation of a thermal signal from deconfined matter
is not excluded. Therefore, we do not attempt any fine tuning of the
codes to reproduce the bulk of data in pp and pA reactions.

\subsection{Physical information encoded in the continuum spectrum}

The thermal dilepton signal with single-electron low-momentum cut-offs
$p_\perp^{\rm min} =$ 2 -- 3 GeV exhibits an approximate plateau in the invariant
mass region 2 GeV $\le M \le 2 p_\perp^{\rm min}$ (see Fig.~4). 
The physical information
 
encoded in the height of the plateau can be estimated by
\begin{equation}
\frac{dN}{dM^2dY}
=
3 \frac{\alpha^2 R_A^2}{4\pi^2} \, F_q \,
(\tau_i\lambda_i^q T_i^3)^2 \,
2 \int_{2p_\perp^{\rm min}/T_i}^\infty dx \,
\left( \frac{8}{x^2} + 1 \right) \, K_3(x) \,
\frac{x - 2 p_\perp^{\rm min}/T_i}{\sqrt{x^2 - (M/T_i)^2}}
\label{approx}
\end{equation}
(cf. Appendix for details of the derivation).
By explicit calculation of this equation and
as seen in Fig.~4, the height of the plateau depends sensitively on the
initial temperature $T_i$ and quark fugacity $\lambda_i^q$.
This dependence can be used to get physical information on the
early and hot stage of the parton matter. In particular, a variation
of the initial temperature within the interval 0.8 -- 1.2 GeV, expected
as possible range of parton matter formed at LHC energies, leads to a
considerable change of the plateau height up to an order of magnitude.
Therefore, once an identification and a measurement of the thermal
dilepton spectrum in the intermediate mass continuum spectrum is possible,
it delivers some implicit information on the initial and very hot
stages of the parton matter. 

\section {Summary} 

In summary we analyze the beam energy dependence of various expected
sources of dileptons in ultrarelativistic heavy-ion collisions. Already
at $\sqrt{s} >$ 20 GeV a copious production of charm gives rise to
a dominant contribution to the dilepton spectrum at intermediate
invariant mass. Taking into account the ALICE detector acceptance we
study also systematically the effect of single-electron transverse
momentum cuts. We find a threshold like behavior of the invariant mass
spectra of dileptons from primary correlated charm and bottom decays and
Drell-Yan yield as well: these sources are suppressed
at $M < M^{\rm min} \approx 2 p_\perp^{\rm min}$ for
$p_\perp^{\rm min} >$ 3 GeV. In contrast to this, the thermal dilepton signal
exhibits a plateau in this region which offers the opportunity to
identify them and to gain information on the initial stages of deconfined
matter at LHC energies.
The same mechanism also works at RHIC energies, however the expected
count rates are too small to make such a strategy feasible.
The complex decay chains of heavy mesons,
in particular open bottom, and the resulting
combinatorial background make an explicit identification of the
''hadronic cocktail'' very desirable to allow a safe
identification of the thermal signal. 

\subsection*{Acknowledgments} 

We thank P. Braun-Munzinger who initiated this investigation.
Stimulating discussions with Z. Lin, R. Vogt, G. Zinovjev are gratefully
acknowledged.
O.P.P. thanks for the warm hospitality of the nuclear theory group
in the Research Center Rossendorf.
The work is supported by BMBF grant 06DR829/1.

\newpage

\section*{Appendix} 

We rewrite the general expression Eq.~(\ref{gen_exp})
for the thermal dilepton spectrum as
\beqn
\frac{dN}{dM^2}
& = &
\frac{\alpha^2 R_A^2}{4\pi^4} F_q
\int d \tau \tau d^2p_+ d^2p_- dy_+ dy_-
\, K_0 \left(\frac{\MT}{T}\right)\lambda_q^2
\nonumber\\
& \times &
\delta\left[M^2- 2 p_{\perp +} p_{\perp -}
\{ \cosh(y_+ - y_-) - \cos(\phi_+ - \phi_-) \} \right]
\eeqn
and arrive after three integrations at
\beqn
\frac{dN}{dM^2}
& = &
\frac{\alpha^2 R_A^2}{4\pi^2}
F_q
\int dY d\MT^2\
d\tau \, \tau \, K_0\left(\frac{\MT}{T(\tau)}\right) \,
\lambda_q^2(\tau) \,
\frac{1}{\pi}
\II(M,\MT,p_\perp^{\rm min})
\label{8}
\eeqn
with
\beqn
\II(M,\MT,p_\perp^{\rm min})= 2\int\limits_{x_1}^{x_2}
dx \frac{1}{\MT}\frac{x^2\ \EllF(\phi,k)
+\MT^2[\EllE(\phi,k)-\EllF(\phi,k)]}{\sqrt{(x^2-\QT^2)(\MT^2-x^2)}}\ ,
\eeqn
where $\EllE$ and $\EllF$ stands for
the incomplete elliptic integrals
\begin{equation}
\EllF(\phi,k)
\equiv
\int\limits_0^{\sin\phi}{\frac{dt}{\sqrt{(1-t^2)(1-k^2t^2)}}},
\quad
\EllE(\phi,k)
\equiv
\int\limits_0^{\sin\phi}{\frac{(1-k^2t^2)\
dt}{\sqrt{(1-t^2)(1-k^2t^2)}}}
\end{equation}
with
$\phi=\arcsin\left(\frac{\Delta(x)}{\QT}\right)$ and
$k=\frac{\QT}{\MT}$, and
$\Delta(x) = \min(x - 2 p_\perp^{\rm min},\QT)$,
$x_1 = \max(2 p_\perp^{\rm min},\QT)$,
$x_2 = \MT$.
In the special case $p_\perp^{\rm min} =$ 0 the expression
$\II(M,\MT,p_\perp^{\rm min})$ becomes $\pi$.
In the small-$M$ region one can approximate
$\Delta(x) \approx M_\perp - 2 p_\perp^{\rm min}$ being independent of
$x$. Then the $x$ integration yields
\beqn
\II(M,\MT,p_\perp^{\rm min})
&\approx&
2\left(\EllE(\pi/2,\sqrt{1-k^2})\EllF(\phi,k)
+ \EllF(\pi/2,\sqrt{1-k^2})\EllE(\phi,k)\right.\nonumber\\
&& \left. -
\EllF(\pi/2,\sqrt{1-k^2})\EllF(\phi,k) \right).
\eeqn
Since for small $M$ also $\QT/\MT\approx 1$ holds, one can use
$
\EllE(\phi,0) = \EllF(\phi,0) = \phi$
and
$\EllE(\phi,1) = \sin\phi$
and finds
\beqn
\II(M,\MT,p_\perp^{\rm min})
\approx
\pi
\frac{M_\perp - 2 p_\perp^{\rm min}}{Q_\perp}.
\label{12}
\eeqn
For $p_\perp$ cuts larger than 2 GeV one gets from
Eqs.~(\ref{8},\,\ref{12})
the approximate expression Eq.~(\ref{approx})
when approximating the $\tau$ integration appropriately
\cite{PRC98}.

\newpage

\begin{figure}[t]
\centering
~\\[-.1cm]
\psfig{file=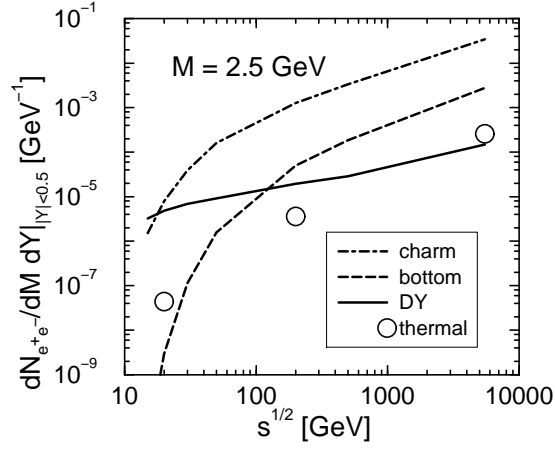,width=6cm,angle=-90}
~\\[.1cm]
\caption{The dependence of dileptons from the lowest-order Drell-Yan
process, and correlated open charm
and bottom decays and the thermal source (only purely
deconfined matter; initial conditions are described in text)
on $s^{1/2}$.
Note that at SPS energies the hadron and a possible mixed phase
(not included here) 
are by far the strongest thermal sources.}
\label{fig.1}
\end{figure}

\vfill

\begin{figure}[b]
\centering
~\\[-.1cm]
\psfig{file=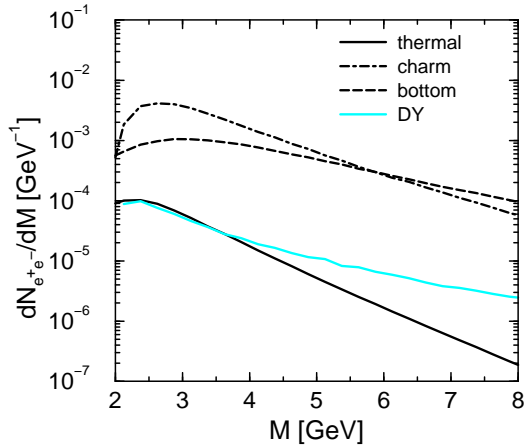,width=6cm,angle=-90}
~\\[.1cm]
\caption{The invariant mass spectra of dileptons from the
Drell-Yan process, and charm and bottom decays, and thermal emission.
The single-electron low transverse momentum cut is
$p_\perp^{\rm min} =$ 1 GeV.}
\label{fig.2}
\end{figure}

\newpage

\begin{figure}[t]
\centering
~\\[-.1cm]
\psfig{file=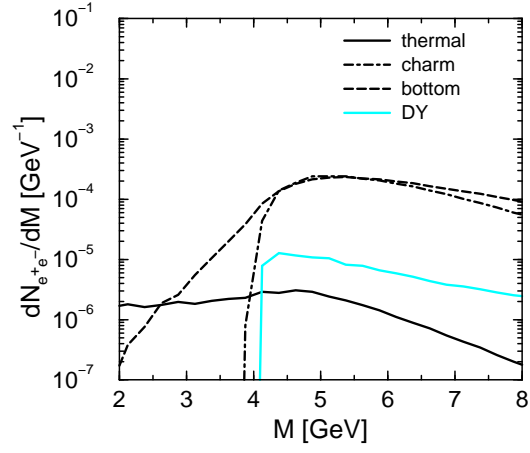,width=6cm,angle=-90}
~\\[.1cm]
\caption{The same as in Fig.~2, but
$p_\perp^{\rm min} =$ 2 GeV.}
\label{fig.3}
\end{figure}

\vfill

\begin{figure}[b]
\centering
~\\[-.1cm]
\psfig{file=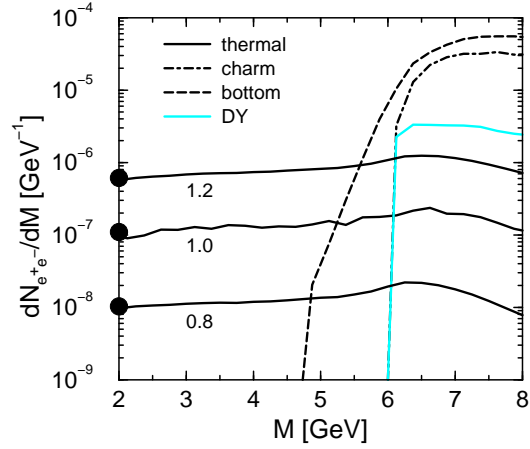,width=6cm,angle=-90}
~\\[.1cm]
\caption{The same as in Fig.~2, but
$p_\perp^{\rm min} =$ 3 GeV and various initial temperatures (in units of GeV)
as indicated by the labels.
The fat dots indicate the estimates of the low-$M$ thermal plateau
according to Eq.~(\protect\ref{approx}).}
\label{fig.4}
\end{figure}

\newpage

\begin{figure}[t]
\centering
~\\[-.1cm]
\psfig{file=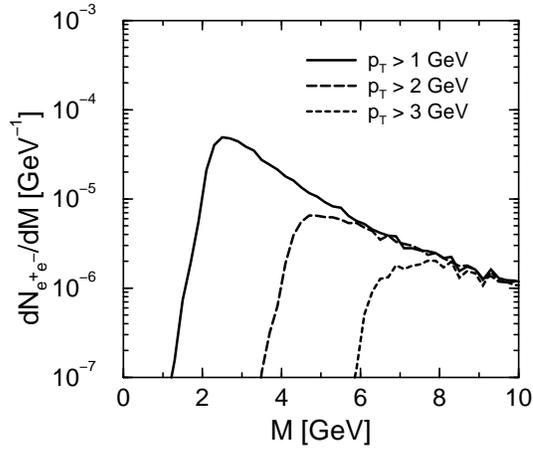,width=6cm,angle=-90}
~\\[.1cm]
\caption{The invariant mass spectra of dileptons from the
Drell-Yan process for $p_\perp^{\rm min} =$
1, 2 and 3 GeV (from left to right). The curves depict results of 
PYTHIA with default switches.
}\label{fig.5}
\end{figure}
\begin{figure}[t]
\centering
~\\[-.1cm]
\psfig{file=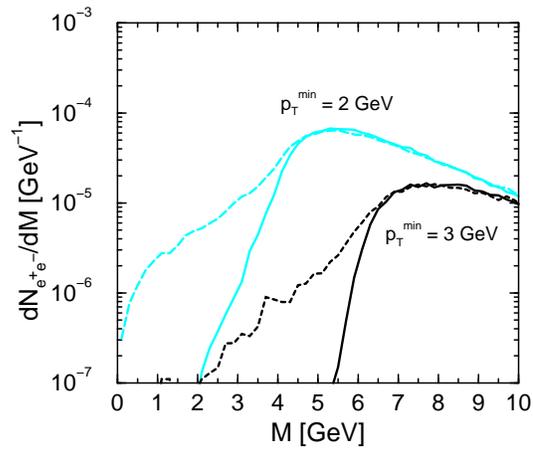,width=6cm,angle=-90}
~\\[.1cm]
\caption{The invariant mass spectra of dileptons from
correlated open bottom meson decays for $p_\perp^{\rm min} =$
2 and 3 GeV. The curves depict results of PYTHIA 
(dashed curves: default switches, 
solid curves: without initial state radiation).
}
\label{fig.6}
\end{figure}

\end{document}